\documentstyle[multicol,prl,aps]{revtex}
\input epsf
\input{epsf.sty}

\begin{document}
\title{Hysteretic behavior of the vortex lattice at the onset of the second peak for HgBa$_2$CuO$%
_{4+\delta }$ superconductor}
\author{D. Stamopoulos and M. Pissas}
\address{Institute of Materials Science, NCSR "Demokritos", 153-10, Aghia Paraskevi, Athens, GREECE.}
\date{\today }
\maketitle

\begin{abstract}
By means of local Hall probe ac and dc permeability measurements we 
investigated the phase diagram of vortex matter for the HgBa$_2$CuO$_{4+\delta }$ 
superconductor in the regime near the critical temperature. The second peak 
line, $H_{\rm sp}$, in contrast to what is usually assumed, doesn't terminate 
at the critical temperature. 
Our local ac permeability measurements revealed pronounced hysteretic behavior 
and thermomagnetic history effects near the onset of the second peak, giving 
evidence for a phase transition of vortex matter from an
 ordered qausilattice state to a disordered glass.

\begin{center}
PACS numbers 74.60.Ge, 74.60.Jg
\end{center}
\end{abstract}

 \begin{multicols}{2}

It is remarkable that after a decade of experimental and theoretical efforts\cite{blatter94,brandt95}
the vortex matter phase diagram of high-$T_c$ superconductors under the presence of random point disorder 
is far from being completely elucidated. 
Soon after the discovery of high-$T_c$ superconductors was proposed  
that point disorder should transform the ideal vortex lattice to a glassy state. This state is characterized  
by energy barriers, $U(j)$ that diverge in the limit of small applied currents, $j$.
Two main phenomenological theories have been proposed to describe this glassy phase.
The first theory is based on the gauge glass model and assumes a complete destruction of the
ideal vortex lattice.\cite{fisher89} The second theory retains the elastic lattice structure at small scales.\cite{blatter94} 
Although different in nature, both theories agreed that point disorder dictates the appearance 
of dislocations, producing the glassy low-temperature phase where the perfect
flux lattice is  completely distorted at large length scales. 

Recently, new theoretical proposals provided a description valid at all scales, demonstrating
that while disorder produces algebraic growth of displacements at short length scales, periodicity
takes over at large scales, resulting in a decay of translational order at most algebraic. 
\cite{Giamarchi,Ertas96,Gingras96,Koshelev98,Kierfield98,Kierfield00} 
One striking prediction is, thus, the existence  of a glass phase that should exhibit Bragg 
diffraction peaks in neutron scattering measurements. \cite{Klein}
This novel vortex state is the so called Bragg glass.\cite{Giamarchi} 
When the field is increased the Bragg glass should undergo a transition into another phase, 
which could be a pinned liquid or another vortex glass. \cite{Giamarchi} 
Such a field-driven transition corresponds to the destruction of the Bragg glass by
proliferation of topological defects, upon raising the field, which is equivalent
to {\it increasing the effective disorder}, which favors dislocations.
Today, the nature of the transition between the two phases, and the exact position of this 
phase boundary on the phase diagram $(H,T)$ are not well understood.  

Between the elastic quasi lattice and the highly disordered vortex glass
there should be a distinct difference, concerning the dependence of
magnetization measurements on the thermomagnetic history, similar as
observed to other disordered systems. 
Such history effects have been observed in transport and magnetic measurements for the low-$T_c$
superconductors, in the region of the conventional peak effect close to the
upper critical field line. \cite{Wordenweber,Ravikumar,Henderson,Bhattacharya,Marley,Ling} 
It has been suggested that in this regime plastic deformations occur in the solid,
leading to a dependence of the hysteretic response on the past history of
the superconductor. 

The situation is rather different for the case of high-$T_c$ superconductors. 
The effect of a second peak (or fishtail peak) in the magnetization loops is still 
of unknown origin, while the simultaneous appearance of double peak structures 
\cite{Zhukov95,Deligiannis97,Rykov99,Zhukov01,Pal01,Sarkar01,Stamopoulos00}  
makes the interpretation more complicated.
Today, there is experimental evidence for the existence of a crossover point
$H_{sp}^{/}$, that lies between the second peak $H_{sp}$ and it's onset point $H_{onset}$,
where the dynamic behavior of the vortex solid changes drastically from elastic
(below $H_{sp}^{/}$) to plastic (above $H_{sp}^{/}$).\cite
{Janossy,Kupfer,Abulafia,Pissas99} 
Furthermore, several experimental works presented evidence that in clean 
YBa$_2$Cu$_3$Cu$_{7-\delta }$ and Bi$_2$Sr$_2$CaCu$_2$O$_{8+\delta}$ crystals the boundary between the
Bragg and vortex glass states is in close proximity to the onset $H_{{\rm onset}}$ of the second
magnetization peak. \cite {Nishizaki98,KhaykovichPRL,Khaykovich97}
In addition, dc magnetization measurements in pure high quality single crystals of 
YBa$_2$Cu$_3$Cu$_{7-\delta }$ revealed the existence of a kink (near the onset of the 
fishtail peak) in the magnetization loops
\cite{Giller99,Radzyner00,Pissas00} and pronounced history dependence 
\cite{Kokkaliaris99,Kokkaliaris00,Zhukov01,Radzyner00} in the regime between the fishtail peak
and it's onset field.

In this paper we report on local dc and ac permeability measurements  
for the HgBa$_2$CuO$_{4+\delta }$ superconductor which displays an intermediate anisotropy 
in comparison to that of YBa$_2$Cu$_3$Cu$_{7-\delta }$ and  Bi$_2$Sr$_2$CaCu$_2$O$_{8+\delta}$. 
With the present study we hope to elucidate the phase diagram of vortex matter 
under the presence of point disorder, near the critical temperature. 
The limited range in the dc field ($H_{{\rm dc}}<1000$ Oe) that can be applied in 
our local Hall magnetometer, enforced us to study a 
disordered single crystal that exhibits a second peak line that terminates in the low field 
regime, close to the critical temperature. \cite{Stamopoulos00}
In our local ac permeability curves we observed hysteretic
behavior at the region between the onset and the second peak (fishtail peak).
Partial loop measurements revealed a pronounced dependence of the ac permeability on the 
thermomagnetic history, in the regime under discussion. 
In addition, this particular single crystal, except for the fishtail peak, displays a third peak 
near the irreversibility line which resembles the conventional peak effect.

The single crystal growth procedure is reported elsewhere.\cite{Pissas97,Pissas98} 
Our single crystal displays a $T_c=89.9$ K with a transition width of $1.5$ K 
and dimensions $600\times 900\times 15$ $\mu m^3$. 
For our local magnetic induction measurements we
used a GaAsIn Hall sensor with an active area of $50\times 50$ $\mu $m$^2$.
The single crystal was placed directly on top of the active area of the Hall sensor. 
The local magnetic induction at the surface of the crystal was measured
using an ac magnetic field ($H_{{\rm ac}}=H_{{\rm 0}}\sin (2\pi ft),$ $f=10$
Hz), under the presence of a dc magnetic field (${\bf H}_{{\rm dc}}\parallel {\bf H}_{{\rm ac}}\parallel c$). 
The real $\mu'=(f/H_{{\rm 0}})\int_0^{1/f}B_z(t)\sin (2\pi ft)dt$ and imaginary 
$\mu''=(f/H_{{\rm 0}})\int_0^{1/f}B_z(t)\cos (2\pi ft)dt$ fundamental
permeabilities are measured by means of two lockin amplifiers. 
ac permeability measurements were performed as a function of the temperature 
(isofield measurements) and also as a function of the applied field (isothermal measurements). 
In addition, we performed local Hall dc magnetization measurements in
order to investigate and clarify the physical mechanisms of the observed
hysteretic behavior. Local dc magnetization measurements were performed by
applying an ac current ($f=10$ Hz) to the Hall sensor and recovering the dc
Hall voltage, due to the dc field, by means of a lockin amplifier. 
Temperature stabilization was better than 20 mK.

\begin{figure}[tbp] \centering%
\centerline{\epsfxsize 7cm \epsfbox{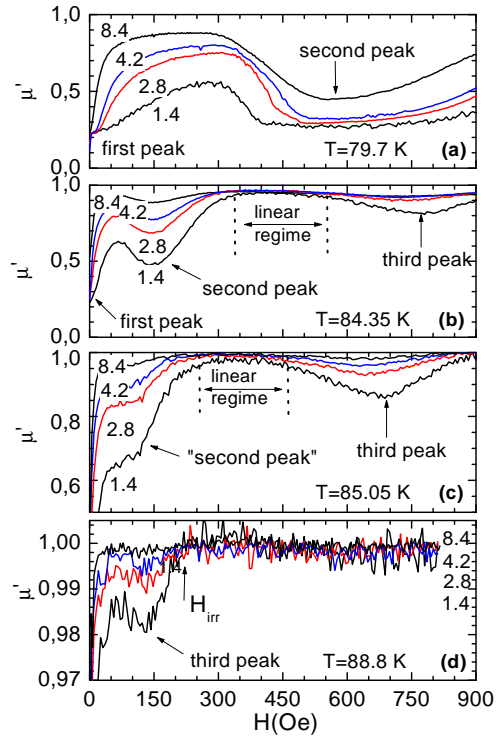}}%
\caption
{
Real part, $\mu'=B'/H_{\rm 0}$, of the local fundamental ac permeability as a function
of dc magnetic field at $T=77.8-89.9$ K for various ac fields, $H_{{\rm 0}}=1.4, 2.8, 4.2, $
and $8.4$ Oe (${\bf H}_{{\rm dc}},{\bf H}_{{\rm ac}}\parallel c$). 
} 
\label{b1}%
\end{figure}

\begin{figure}[tbp] \centering%
\centerline{\epsfxsize 7cm \epsfbox{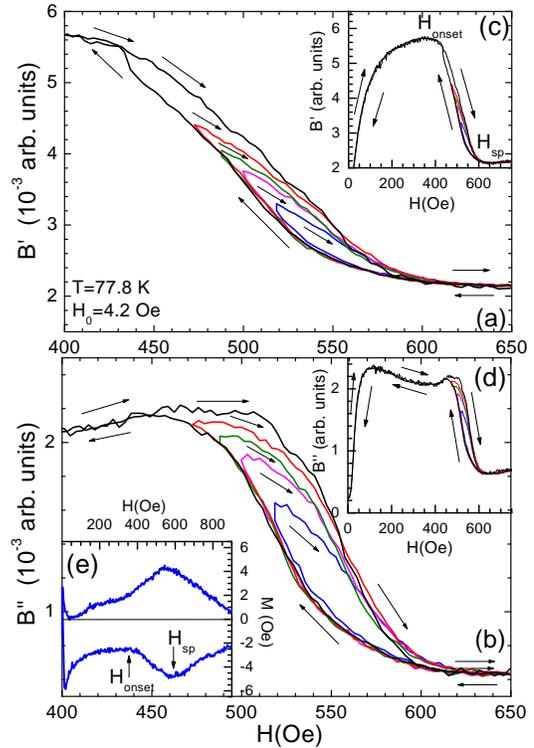}}%
\caption
{
Partial loop measurements of the real, $\mu'$, 
and imaginary, $\mu''$, part of the local 
fundamental ac permeability as a function of dc field for
$H_{{\rm 0}}=4.2$ Oe at $T=77.8$ K (${\bf H}_{{\rm dc}},{\bf H}_{{\rm ac}}\parallel c$).
In the regime between the second peak and it's onset point, pronounced 
thermomagnetic history dependent effects are observed. Insets: In the insets (c),(d) we present the measurement in the whole range.
In inset (e) we present a local dc magnetization loop in order to affirm the results of our ac measurements.
} 
\label{b5}%
\end{figure}

Figure \ref{b1} shows the variation of the real part of the local ac permeability
$\mu'(H)$ as a function of the applied dc field ($H_{{\rm dc}}<1000$ Oe)
for various temperatures and ac fields. The measurement at $T=79.7$ K shows all the characteristic
features for the first and second peaks which are present at the isothermal global magnetization loops.
Initially, $\mu'(H)$ displays a small value due to the finite size of the Hall sensor and it's non zero
distance from the crystal center.
At a particular dc magnetic field, which corresponds to the so called 
first peak field $H_{fp}$, the $\mu'(H)$ starts to increase towards the unit value, which
corresponds to the normal state. Subsequently, instead the $\mu'(H)$ to increase, monotonically, 
up to $\mu'(H)=1$, starts to decrease again, forming  
a local minimum at $H_{\rm sp}$ which obviously corresponds to the second peak as our local and global 
dc magnetization measurements affirmed [see Fig. \ref{b5}(e) below]. \cite {Stamopoulos00}
In the higher temperature regime, in addition to the second peak, a new
local minimum is formed below the irreversibility point [see Figs. \ref{b1}(b)-\ref{b1}(d)]. 
We call this feature as the third peak (referring to a peak in the screening current).
This behavior is recently observed in other high-$T_c$ superconductors.
\cite{Zhukov95,Deligiannis97,Rykov99,Zhukov01,Pal01,Sarkar01,Stamopoulos00}  
In measurements at higher temperatures the height of the local minimum which corresponds to 
the second peak is reduced, and finally, for $T>86$ K, we were not able to detect it.
Contrary to this behavior, the third peak is still evident as we move even up to the critical
temperature. 
Above the irreversibility field, $H_{\rm irr}$ the diamagnetic capability of 
the superconductor becomes zero and all the $\mu'(H)$ curves take the value one.
An important information revealed in those measurements is that, in the region between
the second peak's end point and the third peak's onset the response is
{\it almost linear on the amplitude of the applied ac field}, very close to the normal state value, $\mu'(H)=1$.
This indicates that in this region the vortex system is in a state of very low pinning capability,
resulting in a negligible screening current.

\begin{figure}[tbp] \centering%
\centerline{\epsfxsize 7cm \epsfbox{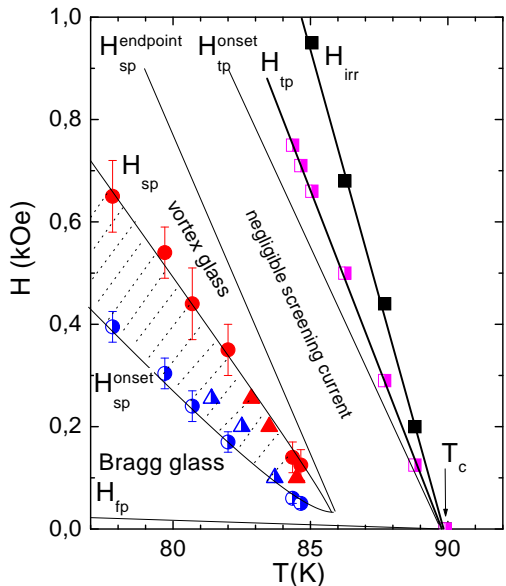}}
\caption
{
The phase diagram of vortex matter close to the critical temperature.
Presented are the onset of second peak (semi-filled circles coming from
isothermal $\mu'(H)$ and semi-filled triangles coming from isofield $\mu'(T)$ measurements), 
the second peak (full circles coming from
isothermal $\mu'(H)$ and full triangles coming from isofield 
$\mu'(T) $ measurements), the third peak (semi-filled squares) and the
irreversibility points (full squares). The second peak line ends at $T{\approx 86}$ K
while the third peak line ends at the critical temperature. The lines are, just, guides
to the eye.
} 
\label{b6}%
\end{figure}

\begin{figure}[tbp] \centering%
\centerline{\epsfxsize 7cm \epsfbox{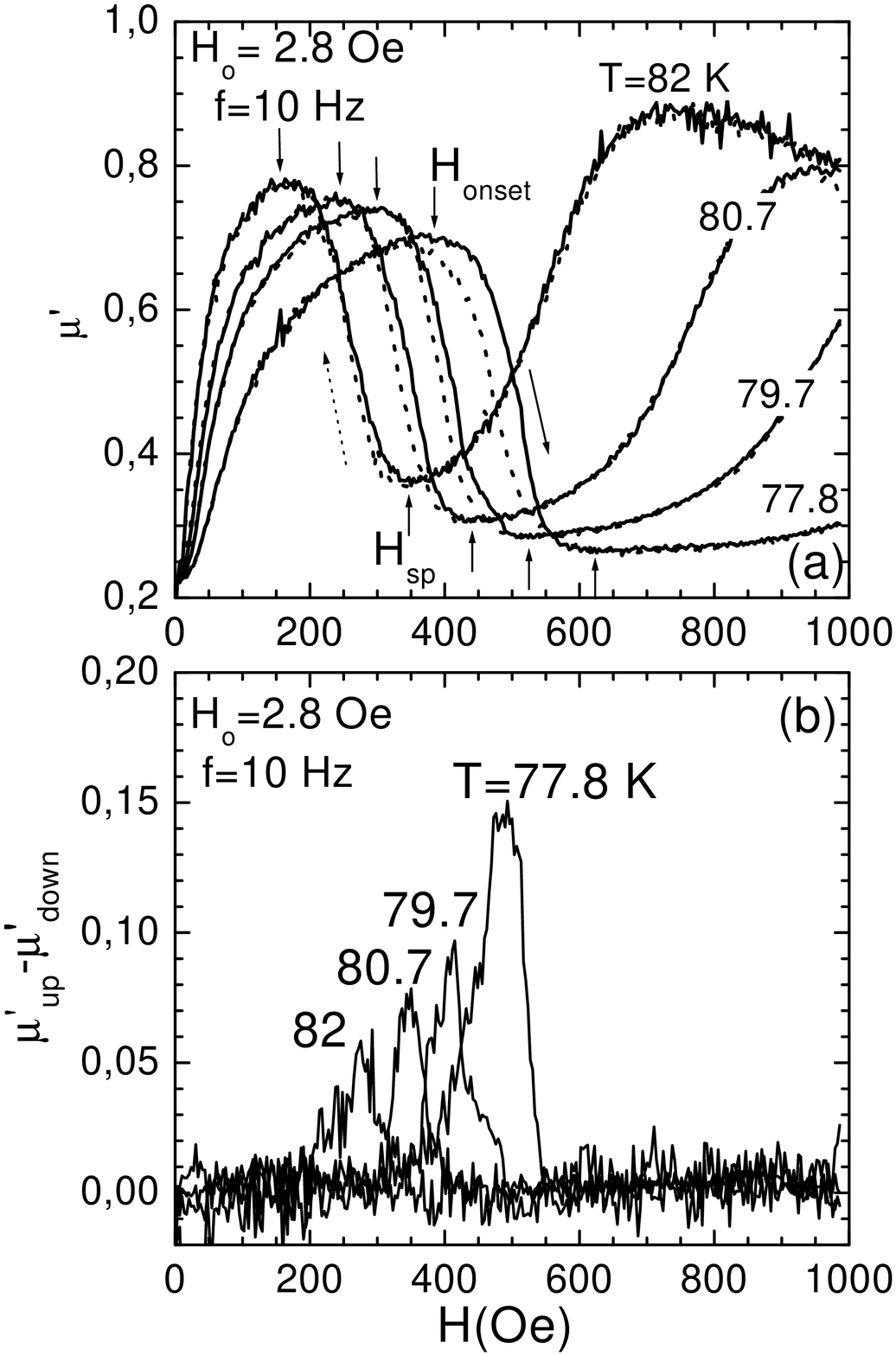}}%
\centerline{\epsfxsize 7cm \epsfbox{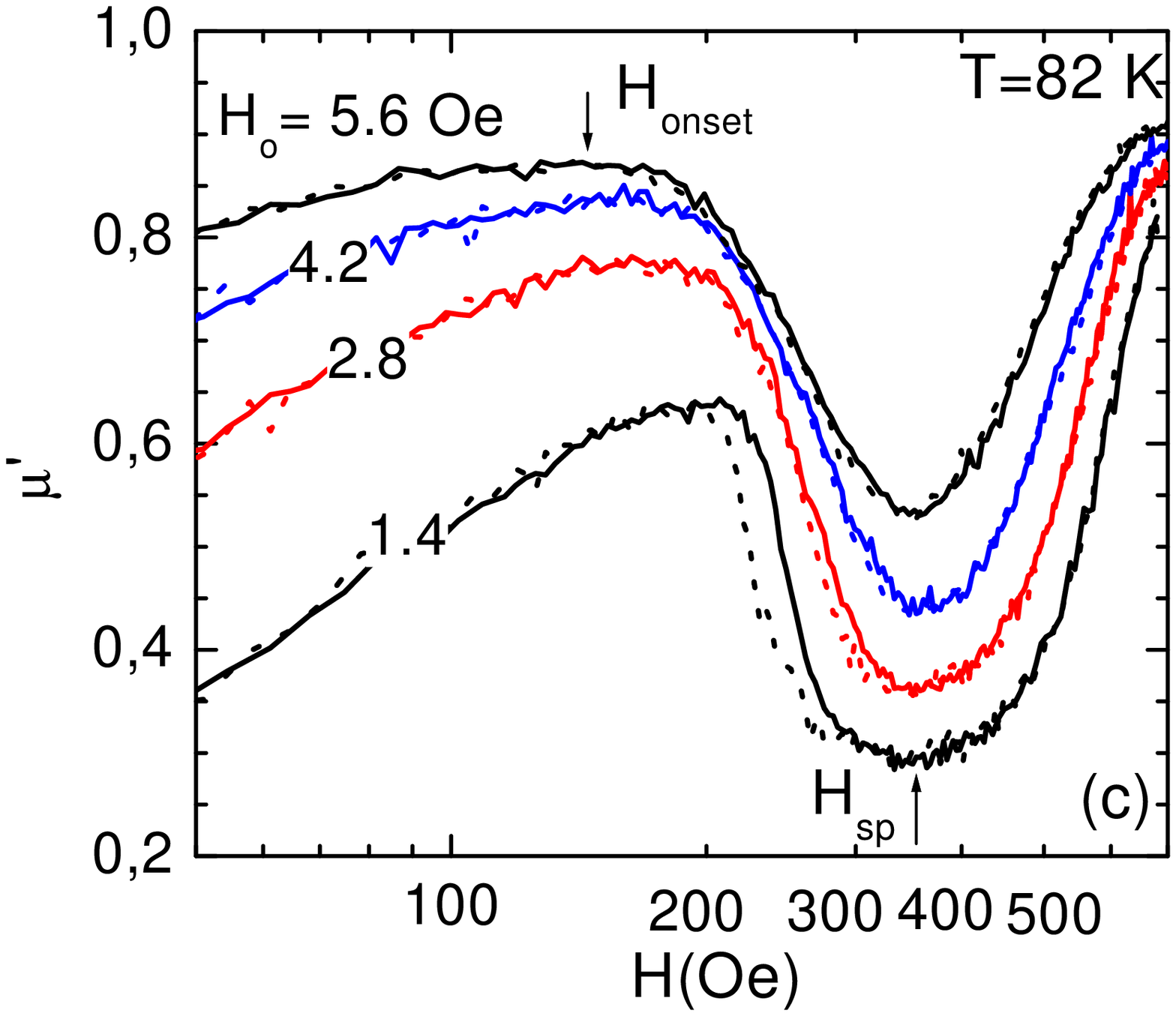}}%
\caption
{
Real part, $\mu'=B'/H_{\rm 0}$, of the local fundamental ac permeability 
as a function of dc field (a) at  constant $H_{{\rm o}}=2.8$ Oe
and  various temperatures $T=77.8, 79.7, 80.7, 82$ K and (c) at constant temperature
$T=82$ K for various ac fields, $H_{{\rm 0}}=1.4, 2.8, 4.2$ and $5.6$ Oe 
(${\bf H}_{{\rm dc}},{\bf H}_{{\rm ac}}\parallel c$). 
The solid (dashed) lines correspond to the increasing
(decreasing) branch. 
(b) The divergence, $\mu'({\rm up})-\mu'({\rm down})$ of the increasing and decreasing branches
for various temperatures is presented. 
} 
\label{b2}%
\end{figure}

In order to fix the previous experimental observations in Fig. \ref{b6} plotted is 
the $(H,T)$ phase diagram close to $T_c$, according to
our local isofield and isothermal ac permeability measurements for the 
HgBa$_2$CuO$_{4+\delta }$ crystal. Depicted are the curves
formed by the onset of second peak, 
by the second peak points (coming from isothermal $\mu'(H)$ and isofield 
$\mu'(T)$ measurements), the third peak and the irreversibility points. 
In the shaded area between the onset
and the second peak lines hysteretic behavior is observed. The points of the second peak's onset may mark the
boundary between the Bragg glass and the disordered glassy state (vide infra). 
Figure \ref{b2}(a) depicts the variation of $\mu'(H)$ as a function 
of ascending and discending the dc magnetic field at various temperatures. 
Although that in general, the decreasing
branch (dashed line) coincides with the increasing one (solid line), we observed
pronounced hysteresis
in the regime between the onset point, $H_{onset}$ and the second peak, $H_{sp}$. 
In Fig. \ref{b2}(b) plotted is the divergence $\mu'({\rm up})-\mu'({\rm down})$ of the data of
Fig. \ref{b2}(a) in order to show clearly the hysteretic behavior. 
We note that the hysteretic peak moves to lower fields for higher temperatures.
We observe that hysteresis is more apparent 
for lower temperatures, as is evident from the reduction of the hysteretic peak as we move to higher temperatures [see Fig. \ref{b2}(b)]. 
As we approach the temperature $T=86$ K, where the second peak line ends, 
the hysteresis is reduced. 
In Fig. \ref{b2}(c) we observe that, at constant temperature, $T=82$ K, the effect is more pronounced
for small amplitudes of the ac field. As we apply higher ac fields the effect is no more evident. 
We observed the same behavior in all the temperature regime, up to the end point $T=86$ K. 
The effect of an applied ac perturbation on the hysteretic behavior, observed in some cases in magnetic 
or transport measurements in high-$T_c$ superconductors, is a
subject of intensive interest.\cite{Henderson,Ling,Valenzuela,Pardo,Fendrich96,Kwok92} 
For small ac fields or small transport currents the driving force acting on vortices is small,
so it may be considered as a small perturbation. In this case, thermodynamic
or dynamic effects characterizing a phase transition may be retained 
(such as hysteresis or a sudden drop in the resistance observed
for example at the melting transition of the vortex lattice). \cite{Charalambous92,Safar92,Fendrich96} 
On the other hand, measurements realized under high transport currents or high ac fields
obscure hysteretic characteristics of a possible underlying phase transition, 
\cite{Kwok92,Fendrich96,Ling} or may dynamically rearrange a disordered
solid state to a more ordered one.\cite{Henderson,Pardo,Valenzuela} 
This is exactly the behavior observed in our measurements. 
For temperatures close to the end point ($T=86$ K) of the second peak line we
were not able to observe hysteretic behavior even for the smallest ac field
we could apply. 

We must note that such small differences can detected only with a sensitive 
local technique, like the one employed in the present work. 
Local magnetic induction measurements by means of microscopic Hall sensors is a
valuable method, because, due to the small size of the active area
of the sensor, the filling factor is unity. So, one can
measure, with high sensitivity, small local changes of the
magnetic induction at the surface of small crystals. In addition,
the achieved high sensitivity (0.01 Oe) permits us to measure
small screening currents such as 10 A/cm$^{2}$ or less. Thereby, we can
estimate with high accuracy, from the onset of diamagnetic
behavior, the irreversibility line or the transition from one
vortex state to another.

Let us now discuss the possible influence of the surface/geometrical barriers 
in our measurements, in order to show that the observed hysteretic behavior 
is directly related to a bulk pinning property of vortex matter.
For the case of a strong non linear current-electric field ($E(J)$) relation
in the glass regime,  one expects a symmetric hysteresis
loop of the global or local irreversible dc magnetization 
(the screening current does not display hysteresis). 
In such case, the fundamental permeability is expected to be independent 
of the measuring path, during descending or ascending the 
external magnetic field, in an isothermal measurement. 
In contrast, for an asymmetric loop of the irreversible 
dc magnetization, hysteresis in the screening current and the fundamental 
permeability should also be observed, during increasing and decreasing the dc field. 
The presence of surface/geometrical barriers generates asymmetry in 
the magnetization loop.
This asymmetry is present in an extensive field range. 
This means that, $\mu'(H)$ is expected to display 
hysteresis in all the range of the loop measurement. 
In our case we may rule out the influence of surface/geometrical barriers,
because the isothermal variation of $\mu'(H)$ displays hysteretic behavior 
{\it only in a particular interval} of the applied dc field. 
This interval is, strictly located between the region of the onset point
and the second magnetization peak.

Another important point revealed in our measurements is that,
at the regime where hysteresis is observed, the decreasing branch is placed
below the increasing one. This means that the high field state of vortex
matter possesses more diamagnetic capability (higher critical current, $J_c$)
than the low field state (zero-field cooling initial condition). 
In the framework of the collective pinning theory \cite{Larkin} the critical
current is related to the characteristic correlation volume $V_c$, over which the vortex solid
is ordered, via the relation $J_c\propto V_c^{-1/2}$. We see that for a more
ordered state (higher collective pinning volume) we have a reduced value for
the critical current. The fact that the high field vortex state exhibits a
higher critical current indicates that is more disordered than the low field one. 
\begin{figure}[tbp] \centering%
\centerline{\epsfxsize 7cm \epsfbox{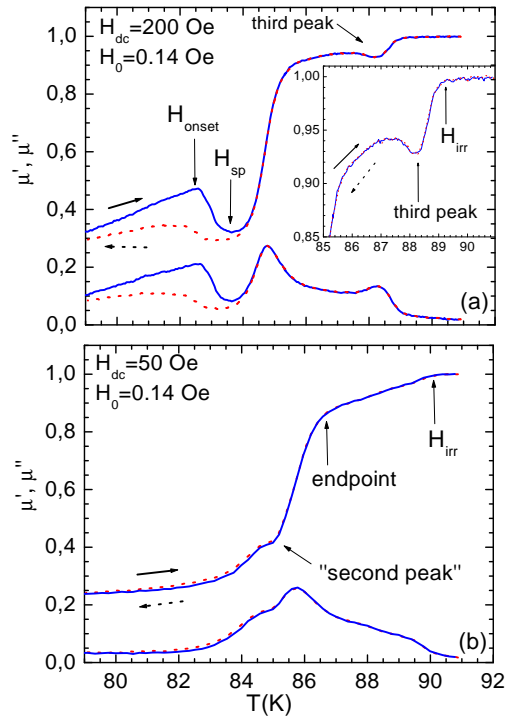}}
\caption
{
Temperature variation of the real, $\mu'$, and imaginary, $\mu''$, part 
of the local permeability, for $H_{\rm dc}=50,200$ Oe and $H_{{\rm 0}}=0.14$ Oe,
for warming (solid lines) above $T_c$, and subsequent cooling (dashed lines)
of the vortex system.
} 
\label{b4}%
\end{figure}%
Pronounced hysteretic behavior could also be observed in isofield ac
permeability measurements as a function of the temperature. 
Fig. \ref{b4} shows  $\mu'(T)$, and $\mu''(T)$ curves for $H_{
{\rm 0}}=0.14$ Oe and $H_{\rm dc}=50$ and 200 Oe. 
The onset, $H_{{\rm onset}}$ and the second peak,
$H_{{\rm sp}}$ points are placed to higher temperatures for
the increasing branches than for the decreasing ones. In the inset we
present the third peak observed just below the irreversibility point. For
high applied ac fields the hysteretic behavior could not be observed, in
agreement to the isothermal $\mu'(H)$, $\mu{''}(H)$ measurements
presented above. 

At this point we must note that, in measurements realised 
under very low dc magnetic fields, the characteristic fingerprint of the fishtail 
peak is transformed to a plateau or a sudden change in the slope of the $\mu'(H)$ 
curve, as is evident for the case of $H_{\rm dc}=50$ Oe. In this low dc field regime
we were not able to detect the unchallengeable structure of the fishtail peak
for all the ac fields that we can apply ($0.01<H_{{\rm 0}}<20$ Oe)

Is worth noticing that the appearance of hysteresis in dynamical measurements 
may be indicative for a possible underlying transition but is not necessarily 
evidential for a phase transition of first order. \cite{Jiang,Geshkenbein} 
In order to understand the physical mechanisms associated to the observed hysteresis 
we performed partial loop measurements. If the observed hysteresis is directly 
related to a first order transition, one would expect that partial hysteresis 
subloops would also be present, due to the finite latent heat and the different 
relaxation rates of the two distinct phases. \cite{Jiang,Geshkenbein}  
Fig. \ref{b5} presents partial loop ac permeability measurements for an ac
field $H_{{\rm 0}}=4.2$ Oe at $T=77.8$ K. 
In the upper and lower panels we
present the in and out of phase local ac magnetic induction signal
respectively. At this measurement the procedure is as follows: starting from
a field above the peak point we perform minor loops by progressively
lowering the minimum value, $H_{min}$ of the applied dc field. Remarkably,
the minor loops do not follow the envelope complete loop. 
We observe that the increasing branch of each minor
loop (corresponding to a lower value of the minimum applied dc field $
H_{min} $) is placed below the corresponding branch for the next value
of $H_{min}$. At the end, arround the onset of the second peak no hysteretic
behavior can be observed. In the insets (c) and (d), we present the complete $B_z^{'}$
(upper panel) and $B_z^{''}$ (lower panel) curves, while in inset (e) we present 
the local dc magnetization loop at $T=77.8$ K, in order to consolidate our experimental results. 
We clearly see that the partial subloops follow exactly the shape of the complete loop, 
without retracing the same curve after reversing the field sweep. 
This is a direct characteristic of a first order transition. In such case we expect
that every fraction of the vortex solid, which transforms in every partial process, 
should follow the same thermomagnetic pattern of the complete transition as in the 
case where the whole vortex system transforms from one state to another. \cite{Jiang,CrabtreeJLTP}
In addition,
the observed thermomagnetic history dependence of the ac response is not
compatible to the conventional critical-state model. This model treats the
critical current, $J_c$ as a single valued function of the magnetic induction $B$
and temperature $T$, while our measurements indicate that $J_c$ depends on the measuring path in the regime between $H_{onset}$ and $H_{sp}$. 
The observed behavior can be understood as follows: as
we expose the system to a lower value of the applied dc field, $H_{min}$
the topological defects remaining in the vortex solid decrease, so as the
critical current is reduced, and the corresponding losses, reflected at the
out of phase signal, increase. Since in the elastic theory, the critical
current $J_c$ is a single valued function of $B$ and $T$, the hardening
effects, such as observed in our measurements, could only be ascribed to {\it plastic
deformations in the vortex solid}. This is in agreement to the behavior observed
in Refs. 20,24-27,32,34,35. 
%
%
%
Finally, we would like to discuss what happens at the region where the fishtail peak 
is terminated. Our local ac permeability measurements as a function  
of dc magnetic field, imply that in the region between the second peak's end point 
and the third peak's onset, the screening current is negligible, as all ready 
discussed above [see Figs. \ref{b1}(b), \ref{b1}(c) and \ref{b2}].
A possible candidate vortex phase, exhibiting such a characteristic behavior ($J_c\sim 0$), 
could be a viscous {\it liquid state} of flux lines. This phase, as we raise the 
temperature or dc field, transforms at the irreversibility point, $H_{irr}$ to 
a {\it gas state} of pancake vortices. \cite{Glazman}
The observed behavior, in this regime, could be also ascribed to a 
weakly pinned {\it solid state}, which in turn exhibits a peak effect 
(third peak) as it melts, under the influence of thermal or quenced disorder. 
Under this point of view, this weakly pinned vortex solid
could be a {\it Bragg glass} state, that reenters through the low field 
regime (below the point at which the fishtail line ends),
or the part of a vortex glass that is placed above the depinning temperature.
As we raise the temperature (or dc field), the solid glass melts by exhibiting
a peak in the screening current (third peak). This is consistent to recent
numerical simulations, revealing that the screening current exhibits a peak both across the Bragg
glass to vortex glass transition and across the melting line. \cite{Otterlo}

In summary, we presented local Hall ac permeability measurements as a
function of the applied dc field (isothermal) and temperature (isofield) for
a HgBa$_2$CuO$_{4+\delta }$ single crystal with $T_c=89.9$ K. 
The second peak line ends a few Kelvins below the critical temperature. This could be 
related to the recently proposed theoretical suggestions \cite{Kierfield98,Gingras96,Giamarchi} 
and experimental evidence \cite{Paltiel00,Ghosh,Banerjee,Pal} for a reentrance behavior of 
the glassy state in the low field regime above the lower critical field line.
At the onset of the second peak we observed pronounced hysteretic behavior,
giving evidence for an underlying first order phase transition
between an almost ordered lattice state, where elastic behavior dominates,
and a disordered glassy state, where plastic deformations are more important.
Recently, experimental evidence has been provided for a first order transition 
between the Bragg glass and the disordered phase in Bi$_{2}$Sr$_{2}$CaCu$_{2}$O$_{8+\delta}$ and 2H-NbSe$_{2}$ 
single crystals.\cite{Beek00,Giller00,Gaifullin00,Avraham,Paltiel00} 
We hope that our results will assist the effort for the investigation of the nature 
of the order-disorder transition between the onset and the second peak.

\begin{acknowledgments}
This work was supported from the Greek Secretariat for Research and Technology through 
the PENED program (99ED186) and Dimoerevna program (642).  
\end{acknowledgments}

\end{multicols}

\end{document}